\title{Optimising optical tweezers experiments for magnetic resonance sensing with nanodiamonds} 
\documentclass[journal=jacsat,manuscript=article]{achemso}

\usepackage{chemformula} 
\usepackage[T1]{fontenc}

\usepackage{graphicx}
\usepackage{dcolumn}
\usepackage{bm}
\usepackage{color}
\usepackage{ulem}
\usepackage{natbib}

\definecolor{lwr}{RGB}{0, 128, 128}
\definecolor{pjr}{RGB}{255, 67, 164}
\definecolor{ds}{RGB}{128, 0, 164}
\definecolor{aaw}{RGB}{93, 49, 63}

	\author{L.W. Russell}
	\affiliation{School of Physics, The University of New South Wales, Sydney, NSW 2052, Australia}
	\author{E.C. Dossetor}
	\affiliation{School of Physics, The University of New South Wales, Sydney, NSW 2052, Australia}
	\author{A.A. Wood}
	\affiliation{School of Physics, The University of Melbourne, Parkville, VIC 3010, Australia		 	}
	\author{D.A. Simpson}
	\affiliation{School of Physics, The University of Melbourne, Parkville, VIC 3010, Australia		 	}%
	\author{P.J. Reece}
	\email{p.reece@unsw.edu.au}
	\affiliation{School of Physics, The University of New South Wales, Sydney, NSW 2052, Australia}

	\title{Optimising optical tweezers experiments for magnetic resonance sensing with nanodiamonds} 	
	
\begin{document}

\begin{abstract}
In this article we explore the requirements for enabling high quality optically detected magnetic resonance (ODMR) spectroscopy in a conventional gradient force optical tweezers using nanodiamonds containing nitrogen-vacancy (NV\textsuperscript{--}) centres. We find that modulation of the infrared (1064 nm) trapping laser during spectroscopic measurements substantially improves the ODMR contrast compared with continuous wave trapping. The work is significant as it allows trapping and quantum sensing protocols to be performed in conditions where signal contrast is substantially reduced. We demonstrate the utility of the technique by resolving NV\textsuperscript{--} spin projections within the ODMR spectrum. Manipulating the orientation of the nanodiamond via the trapping laser polarisation, we observe changes in spectral features. Theoretical modelling then allows us to infer the crystallographic orientation of the NV\textsuperscript{--}. This is an essential capability for future magnetic sensing applications of optically trapped nanodiamonds.
\end{abstract}

\maketitle


%


\section{Introduction}

The nitrogen-vacancy (NV\textsuperscript{--}) defect centre in diamond represents one the most promising emerging technologies for high precision nanoscale sensing. Spin preserving optical transitions between the ground and excited state in the NV\textsuperscript{--} centre have different relaxation pathways depending on the initial spin state, allowing them to be distinguished through optical spectroscopy. The application of microwave fields at the transition frequency between non-degenerate m\textsubscript{s} = 0 and $\pm1$ states can be used to control the spin state of the NV\textsuperscript{--}  and can be used for magnetic resonance measurements. The sensitivity of the optical transitions and underlying spin states to environmental factors, such as local DC and AC magnetic fields, electric fields, strain, and temperature, makes it an excellent candidate for multi-modal sensing.

The unique opportunity for NV\textsuperscript{--} centres is that high spatial resolution and sensitivity combined with a relatively simple measurement scheme makes new studies possible where traditional technologies (e.g. SQUID magnetometers) are incompatible. Already the application of NV\textsuperscript{--} sensing protocols have been used to map carrier transport in 2D condensed matter systems (e.g. graphene) \cite{Tetienne2017}, measure the dynamics of single spins \cite{Lovchinsky2017,Shi2015}, perform NMR spectroscopy on micro-litre fluid volumes \cite{Smits2019}, and monitor neuronal behaviour at the single nerve level \cite{Barry2016}. One of the limitations of precision NV\textsuperscript{--} sensing in bulk diamond is that it is mainly limited to studying two-dimensional systems. Because of the small interaction volume, typical experimental configurations rely on preparing target systems directly atop high quality diamond substrates with NV\textsuperscript{--} centres located near the surface. However, in many cases, sensing away from a substrate would be advantageous.

Nanodiamonds (NDs) with dimensions of approximately 100 nm partially fulfil this requirement. Typically prepared by irradiation of ball-mill processed high-temperature high-pressure (HPHT) diamond, NDs support magneto-optically active NV\textsuperscript{--} centres and can be used in a wide range of experimental situations. Previously, such NDs have been injected into a living cell and used as a nanoscale temperature probe \cite{Kucsko2013}. Challenges associated with using NDs are that the spin properties of the NV\textsuperscript{--} centres can be quite varied between individual nanodiamonds (influenced by surface states and the presence of other paramagnetic defects) and there is little control over how the nanodiamonds are deployed or directed to points of interest. Optical tweezers allow for the controlled positioning and orientation of nanoparticles inside a three-dimensional microscopic environment, including within living cells \cite{Norregaard2014}. This is particularly important in considering the highly localised nature of the sensing region. Trapping avoids the reliance on passive diffusion to deliver NDs to areas of interest whilst retaining the ability probe volumes in three dimensions, which is lost when using NV\textsuperscript{--} centres embedded in substrates. 

In the area of ND trapping and sensing Geiselmann \textit{et al.} \cite{Geiselmann2013} performed ODMR measurements on single NDs containing a single NV\textsuperscript{--} centre. They were able to show both orientation control as well as modulation of the fluorescence signal through interactions with surrounding media. At this time, Horowitz \textit{et al.} \cite{Horowitz2012} similarly showed ODMR could be obtained from ensembles of NDs containing thousands of NV\textsuperscript{--} centres. They were able to measure spectral changes in the presence of an external magnetic field, thereby demonstrating the first magnetometry based sensing on trapped NDs. Work by Juan \textit{et al.}  \cite{Juan2017} extended this work by looking at the effects of trapping of single NDs in resonance with the NV\textsuperscript{--} optical transition. In parallel there has been a heavy focus on quantum sensing combined with related techniques of vacuum trapping and levitation \cite{Neukirch2013,Pettit2017}. 

One important presumption in all of these discussions is that the trapping techniques do not perturb the behaviour of the sensor. A na\"{i}ve and incorrect assumption is that the trapping photon energy (typically in the infrared) is too small to excite optical transition from the NV. In fact, the influence of the infrared (IR) illumination on the NV\textsuperscript{--} photoluminescence is substantial and has been found by multiple studies to cause significant quenching, hence reducing contrast in optical measurements \cite{Geiselmann2013b}. This is a topic that is of general interest to the quantum sensing community and there are a number of conflicting results with no strong consensus on the exact mechanism \cite{Ji2016}.

In our recent work we have also found that the infrared trap can dramatically modify the spin behaviour as measured through spin-lattice (T\textsubscript{1}) relaxation \cite{Russell2018}. This effect was so strong that at any moderate trapping powers the spin polarisation properties were completely lost. Through judicious choice of measurement protocol we are able to recover the spin polarisation and enable T\textsubscript{1} based sensing without compromising the optical trapping performance. With the problem of the photoluminescence quenching and depolarisation addressed we are able to start to explore the full capabilities of quantum sensing for optically trapped NDs.

In this work we extend our method of trap modulation to enable high quality pulsed ODMR spectroscopy independent of the applied trapping intensity. Starting with continuous wave trapping experiments we show how there is a rapid loss of ODMR contrast with increased trapping power. By modulating the trapping laser such that the excitation and read-out of the magnetic resonance is done in the absence of the infrared laser, the signal contrast is fully restored. Operating in this mode we are able to achieve DC magnetic field sensitivity surpassing that of Horowitz \textit{et al.} by an order of magnitude on individually trapped NDs. By mitigating the infrared induced quenching effects we are also able to perform dual mode sensing on sub-50 nm diamonds where fluorescence intensity is intrinsically lower. Finally, with increasing applied magnetic field we observe peak splitting of ODMR signals consistent with the four different projections of a single ND containing an ensemble of NV\textsuperscript{--} relative to a fixed crystallographic orientation. Rotating the trapping laser polarisation through $90^\circ$ we observe changes to the ODMR spectrum consistent with predicted Zeeman splitting with a corresponding rotation of the NV axis. These results are supported by theoretical modelling of the NV Hamiltonian in the presence of an off-axis magnetic field. This indicates that the single ND  maintains a specific orientation with the trap and that can be manipulated for vector magnetometry. 

\section{Experimental Arrangement}

\begin{figure*}[ht!]
\begin{center}
\includegraphics[width=\linewidth]{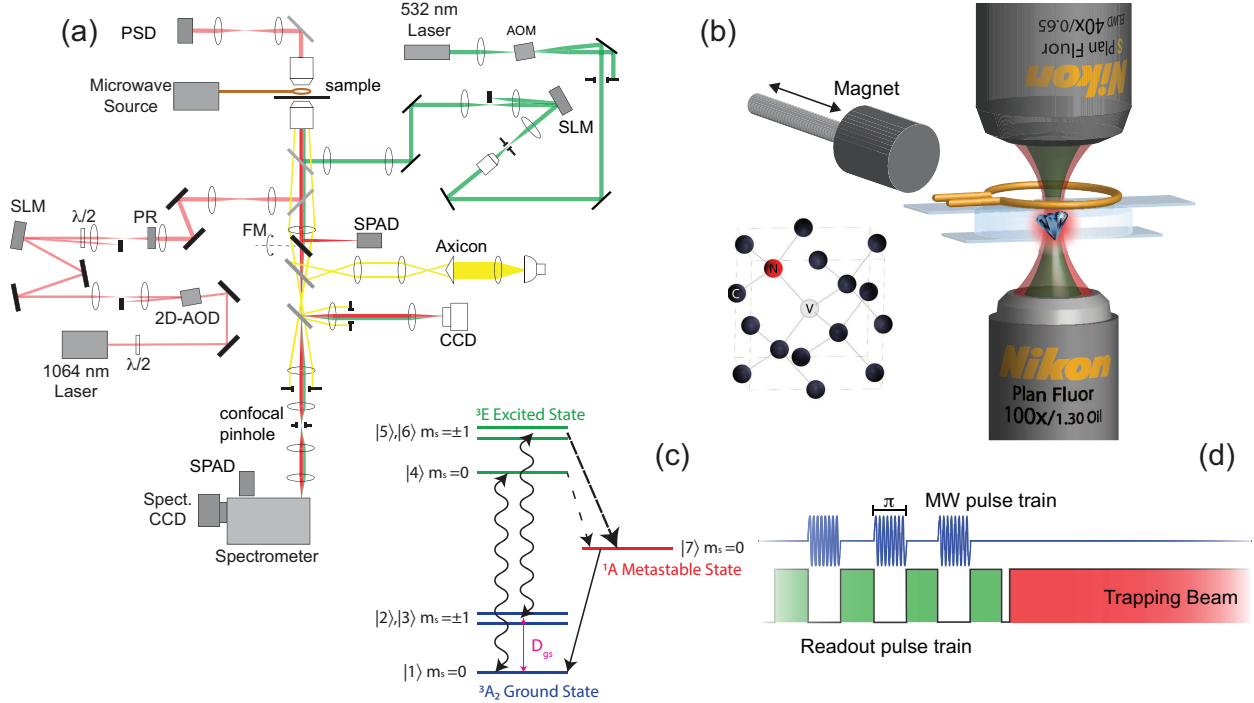}
\end{center}
\caption{ a) Schematic of the optical microscope used for the study: it includes an infrared (1064 nm) trapping laser, green excitation laser (532 nm) and a confocal imaging / spectroscopy optics. b) ND samples are trapped inside a thin fluidic chamber bounded by two coverslips. An external loop antenna is used to apply microwave pulses and an external DC magnetic field is supplied by a rare earth magnet mounted on a translation rod. The trapped NDs contained several hundred NV\textsuperscript{--} defect centres (insert) which emit bright stable fluorescence in the red/near-infrared spectral band. c) Energy level diagram corresponding to the optical transitions of the NV\textsuperscript{--} defect centre in diamond. The ground triplet spin states have zero-field splitting of 2.87 GHz between the m\textsubscript{s} $ = $ 0 and m\textsubscript{s} $ = \pm $ 1 states. A meta-stable singlet state provides a non-radiative pathway for polarising to the m\textsubscript{s} $ = $ 0 NV spin ground state.  d) Pulsed ODMR experiments are performed during the off state of the modulated trapping laser. The measurement sequence includes a green laser polarising/readout pulse followed by a microwave $\pi$-pulse. The microwave frequency is swept through the NV\textsuperscript{--} spin resonance.}\label{fig:1}
\end{figure*}

The experimental apparatus used for this study is depicted in Fig. \ref{fig:1}a. The gradient force optical tweezers consists of a linearly polarised Nd:YAG infrared laser (Laser Quantum, IR Ventus) with $ \lambda =  $ 1064 nm, TEM\textsubscript{00}, focused through a 1.3 NA oil-immersion microscope objective (Nikon CFI Plan Fluor 100$ \times $) to a diffraction limited spot in the front focal plane. Modulation and positioning of the beam is achieved using a two-axis acousto-optic deflector (2D-AOD, Gooch \& Housego 45035 AOBD) which is controlled with two digitally addressable digital frequency synthesisers (DFS, Gooch \& Housego, 64020-200-2ADMDFS-A). A spatial light modulator (SLM1, Hamamatsu LCOS-SLM x10468-03) is incorporated to correct for aberrations in the beam profile. A liquid crystal polarisation rotator (PR, Meadowlark Optics, LPR-100-1064) inserted in the beam path is used to control the orientation of the linear polarisation state at the trapping focus. The Brownian dynamics of the trapped ND are measured using back focal plane interferometry of the forward scattered infrared light \cite{Gittes1998}, which is collected through a condenser lens (Nikon CFI Plan Fluor ELWD 40X) projected onto a position sensitive detector (PSD, Pacific Sensor DL16-7PCBA).

The optical excitation is provided by a green optically pumped semiconductor laser (Coherent Verdi G7) with $ \lambda = $ 532 nm, which is modulated using an acousto-optic modulator (AOM, Gooch \& Housego 3110-120) with power controlled using a half-wave plate and polarising beamsplitter. The laser is coupled to the microscope column using a dichroic mirror and focused to the trapping site such that the focus is coincident with the trap position; fine position control is achieved via a second spatial light modulator (SLM2, Hamamatsu LCOS-SLM x10468-03). Spectrally resolved photoluminescence is collected through the focusing objective and relayed, via a 150$\,\mu$m confocal aperture to a spectrometer (Princeton Instruments, Acton SP2300) fitted with a PIXIS 256 spectroscopic CCD. For ODMR measurements, emission was imaged on to to a single photon counting avalanche photo-diode (APD, PicoQuant  $\tau$-SPAD), via a flipper mirror and band pass filter (Semrock FF02-675/67-25) placed directly after the microscope column. The small detector dimensions 150$\,\mu$m acted as a confocal aperture to select out emission from the trapped ND.

In Fig. \ref{fig:1}b a graphic is presented depicting the set-up near the focus. The sample chamber consists a fluid cell containing 5 $\mu$L of aqueous solution NDs bounded by two coverslips mounted on a custom made slide holder. The sample is mounted on the microscope stage and a loop antenna (diam. $\sim$5 mm) is brought close to the top surface between the sample and the condenser lens. The antenna is connected to a Rohde \& Schwarz  vector signal generator (SMBV100A) and a Mini Circuits (ZHL-25W-63+) power amplifier, which is used to apply a microwave field in the frequency range of 2.87 GHz, which corresponds to zero-field splitting of the m\textsubscript{s} $ = $ 0 and m\textsubscript{s} $ = \pm $ 1 states (Fig. \ref{fig:1}c). An external magnetic field was provided by a permanent rare-earth magnet (cylindrical with 19 mm diameter, 28.2 mm length), mounted on a long travel translation stage. The magnet was oriented such that its cylindrical axis at an angle of inclination of 23$^\circ$ with respect to the imaging plane and azimuthal angle of 13$^\circ$ with the in-plane y-axis. This orientation was chosen to give the maximum in-plane field without obstructing the sample stage. The field strength at the sample position is characterised with a gaussmeter (Lake Shore Model 410) with a transverse Hall probe (Lake Shore MST-9P04-410).
 
In order to co-ordinate the control of trapping, excitation and detection, a high speed programmable pulse generator (SpinCore PulseBlasterESR-PRO) is used. This is able to trigger the photon collection as well as modulate the trapping and excitation lasers via the 2D-AOD \& AOM respectively. For ODMR, binning of the photon counts is performed with the counter and timing functionalities of a data acquisition card (DAQ, National Instruments PCI-6251), which also triggers the frequency sweep of the signal generator. Complementary spin relaxometry measurements are performed using time correlated single photon detection and utilises a multi-channel scalar (MCS, FastComTec MCS64A) that is also triggered by the pulse generator.

\section{Results and Discussion}

\begin{figure}[h!]
\begin{center}
\includegraphics[width=\linewidth]{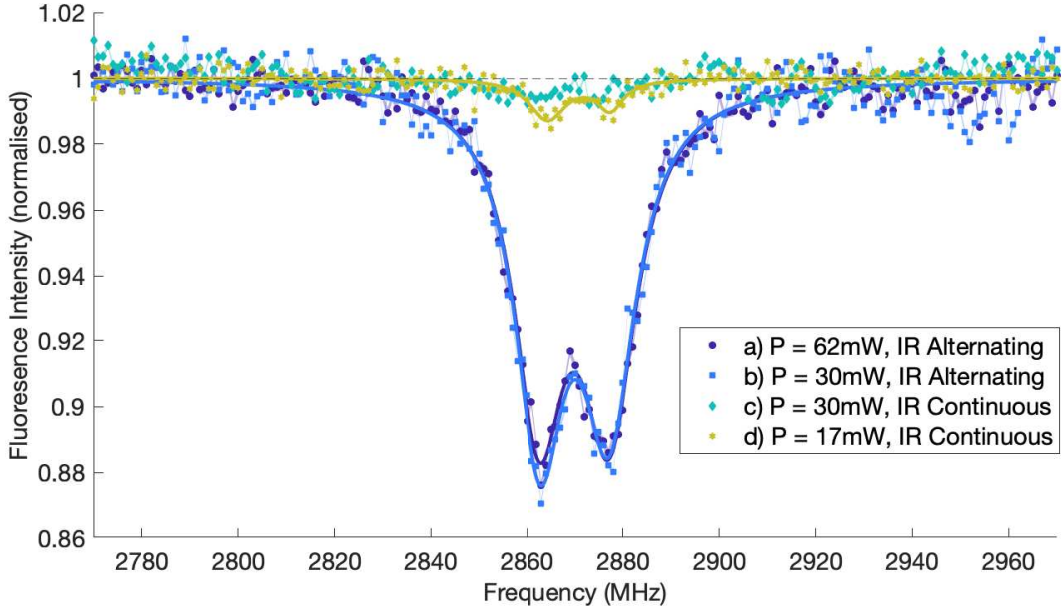}
\end{center}
\caption{ODMR of a single trapped 100 nm nanodiamond with varying trapping laser settings. a-b) ODMR with the trapping laser temporarily switched off during the ODMR pulse sequence, i.e. `IR Alternating', performed at moderate trapping powers (62 and 30 mW respectively). For this case ODMR contrast and linewidth are independent of trapping power. c-d) ODMR with the IR left on during the ODMR pulse sequence, i.e. `IR Continuous', performed at comparable trapping powers. For this case ODMR contrast is strongly suppressed, and appears to be inversely proportional to trapping power.}\label{fig:2}
\end{figure}

In the first set of experiments we explore the influence of the infrared trapping laser on the pulsed ODMR signal from single trapped NDs containing several hundred defect centres and with dimensions of approximately 100 nm (FND Biotech, brFND-100). A schematic of the spin dependent optical transitions in the NV\textsuperscript{--} centre is presented in Fig. \ref{fig:1}c \cite{Doherty2013} and corresponding pulse sequence is shown in  Fig. \ref{fig:1}d. In this scheme, a green excitation laser pulse of 300 ns is used to drive the NV\textsuperscript{--} into the m\textsubscript{s}$ = 0$ triplet ground state by optically cycling the NV\textsuperscript{--} spin through the meta-stable singlet state, $^{1}\textrm{A}$, via the excited state $^{3}\textrm{E}$. Upon spin-polarising the NV\textsuperscript{--}, a microwave field is applied to the ND; if the microwave frequency corresponds to the energy separation between the m\textsubscript{s} $ =  $ 0 and $\pm 1$ states the microwave field will drive the NV spin into the m\textsubscript{s} $ = \pm $ 1 state. Under these resonance conditions the subsequent excitation laser pulse will yield a reduced emission intensity. The microwave $ \pi $-pulse duration corresponds to a 400\,ns, estimated from Rabi measurements.

In Fig. \ref{fig:2} we present plots of the normalised fluorescence intensity versus applied microwave frequency for different trapping powers under two different conditions: (i) for the plots labelled as `IR Continuous', the trapping laser is left on during the ODMR pulse sequence; and (ii) for plots labelled `IR Alternating' the trapping laser is temporarily switched off while the optical signal is being measured. We observe that the under continuous illumination with moderate trapping powers (30 mW) the ODMR signal is strongly suppressed with no apparent contrast at the expected microwave resonance frequency at 2.87 GHz. When the trapping power is decreased (to 17 mW), the resonance dip begins to emerge, but with a significantly reduced contrast. Further reduction in trapping power leads to a trap with insufficient strength to localise the ND for the duration of the experiment. In comparison, when the IR trapping beam is switched off during optical measurements, we see a clear ODMR spectrum  with signal contrast of $ \sim $10\% a line width of $ \sim $15 MHz per peak and with characteristic splitting of the peaks, which have been alternately assigned to strain effects \cite{Gruber1997} in the diamond lattice, or Stark splitting due from the near-lying charge states \cite{Manson2018}. Further, the ODMR contrast is independent of the trapping power in the IR alternating mode and as can be seen by comparing the  ODMR from two different trapping powers of 62 and 30 mW. This means that we may arbitrarily set the trapping power to optimise the trapping conditions. We also note that the line width of the ODMR is comparable to that of other spectroscopy experiments using similar NDs  \cite{Simpson2017} and are significantly narrower than previously reported ODMR from trapped ND ensembles \cite{Horowitz2012}. 

The sensitivity of the magnetometer to DC magnetic fields is determined by the following expression (from: \cite{Dreau2011}):

\begin{equation}
	\eta_{dc}=\frac{4}{3\sqrt{3}}\frac{h\delta}{ g_{_{NV}}\mu_{_B} R \sqrt{N}}
\end{equation}

Where $ \delta $ is the full width at half maximum of an ODMR peak,  $g_{_{NV}}=$ 2.0028 is the g-factor of the NV centre, $ \mu_{_B} $ is the Bohr magneton, $ R $ is the ODMR contrast and $ N = $ 2140 kCts/s is the fluorescence count rate over the duration of the measurement, the best result achieved in this paper (Dataset Fig. \ref{fig:2}a) yields 5.5 $\mu \mathrm{T}/\sqrt{\mathrm{Hz}}$. This value surpasses by an order of magnitude that found by \citeauthor{Horowitz2012} on an ensemble of trapped NDs. We note that the high count rates and preserved signal contrast using our modulation technique means that acquisition times for ODMR measurements can be quite rapid. For example, the high quality spectrum of Fig.\,\ref{fig:2}a was recorded in 205 s, and lower tolerances on noise can yield spectrally resolved signals in less than a minute. Fast acquisition will be critical for future studies that involve measuring dynamic changes to the environment. 

The method of trap switching here has been used extensively in the past and is referred to as time sharing / time division multiplexing. It is used as a way of creating multiple virtual traps by switching the trapping position rapidly between different spatial locations \cite{Ruh:11}. Each time the trapping laser is cycled through a specific location, the object will experience a restoring optical force. The switching rate in time-division multiplexing must be sufficient to counter the diffusion of the particle, which is in turn governed by the Stokes' drag. As optical forces are a time-averaged effect, the trap stiffness in the alternating trap is then reduced by the proportion of the dwell time within a cycle \cite{Ren2010}. Thus for our experiments the trap stiffness is reduced proportionally to the IR beam duty cycle, which is itself set by considering the timing capabilities of the 2D-AOD, as well as the requirements of the measurement sequence and the Brownian dynamics of the ND.

\begin{figure}[h!]
\begin{center}
\includegraphics[width=\linewidth]{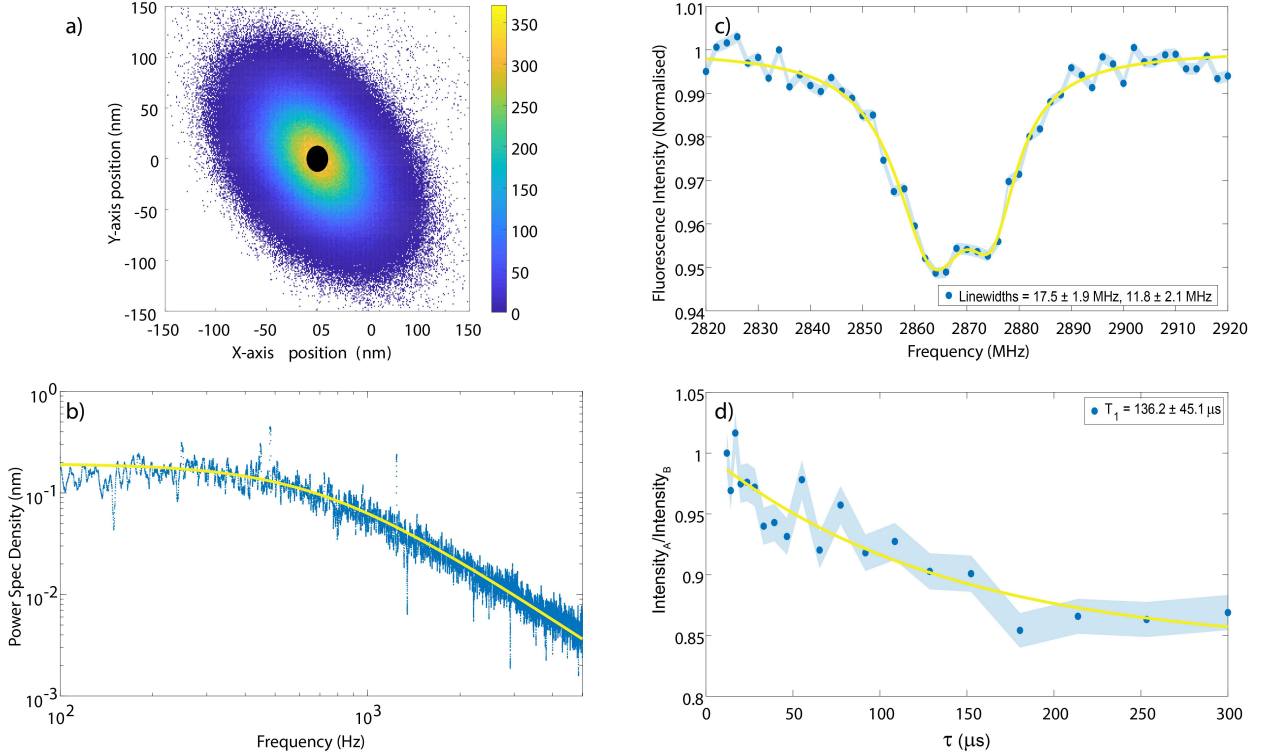}
\end{center}
\caption{Demonstration of multi-modal sensing capabilities from a single trapped sub-50 nm nanodiamond. a) The 2D lateral position histogram of the  stochastic motion of the trapped nanodiamond. Using the Equipartition theorem we estimate the trap stiffness to be 0.76 pN/$\mu$m in the X direction and 0.93  pN/$\mu$m in the Y direction for a 30 mW trap. b) A log-log plot of the power spectrum of particle trajectory in the range of 0.1-5 kHz, together with a Lorentz fit. Assuming a Stokes' drag relation we estimate the particle size to be of the order of 21 nm in X and 26 nm in Y (indicated by the dimensions of the black ellipse in (a)). c) Pulsed ODMR and d) Spin-lattice (T\textsubscript{1}) relaxation from the same ND showing dual sensing capability of the system.}\label{fig:3}
\end{figure}

The sensitivity of NV\textsuperscript{--} based sensors depends on the standoff distance between the NV\textsuperscript{--} centres and measurement targets \cite{Maletinsky2012}. Therefore the size of the ND plays a key role in determining the overall measurement sensitivity. In some cases, e.g. spin relaxometry, the interaction between the NV spins and  surrounding spins is highly localised and as such NV centres lying deep within the ND are not influenced by the environmental spins and simply contribute to lowering the contrast of the signal. In other cases the size of the ND simply defines the volume of the sensing element. For these reasons it is important to minimise the size of the NDs. This presents a problem for optical tweezers experiments as smaller NDs generate highly reduced optical forces and require high trapping powers to confine the particles against Brownian motion. 

In order to test the ability of our approach for trapping and sensing of small particles we perform single particle measurements on commercial NDs that have a nominal size of 35 nm (FND Biotech, brFND-35). In Fig. \ref{fig:3} we present both nanoscale position measurements (a,b) and NV spin based measurements (c,d) performed on the same trapped sub-50 nm ND. a) shows the 2D lateral position histogram showing that the diamond is largely confined to a $ \sim $200 nm elliptical region in the lateral plane, further the inferred lateral size of this ND ($ \sim  $21 nm in X, $ \sim  $26 nm in Y) is overlayed as a black ellipse. The size estimate of the ND is based on an analysis of its Brownian motion, by fitting the Lorentzian position power spectrum (Fig. \ref{fig:3}b) and the Gaussian position histogram and then comparing the extracted trap stiffness values we are able to extract a size parameters \cite{Berg-Sorensen2004}. Nanoparticles typically orient themselves with their long axis either along the direction of the laser's polarisation \cite{Andres-Arroyo2016}, or along the propagation direction \cite{Andres-Arroyo2018}, given the small dimensions estimated in the lateral directions, and the overall stability of its trapping, we suspect this ND is elongated in the Z-direction. 

\begin{figure}[h!]
\begin{center}
\includegraphics[width=\linewidth]{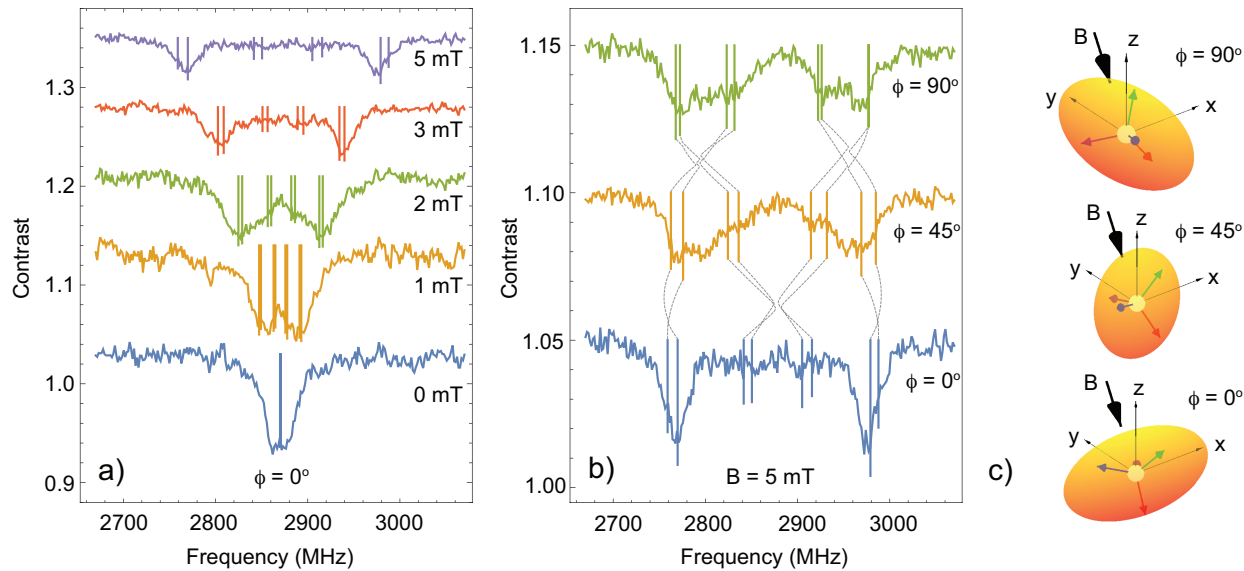}
\end{center}
\caption{a) Demonstration of the magnetic field dependence of the ODMR spectrum for a single trapped 100\,nm ND. As the field is increased from 0 to 5 mT the resonance peak splits into multiple peaks corresponding to the Zeeman splitting of the degenerate m\textsubscript{s} $ = \pm $ 1 and the four possible NV axis projections for a fixed crystal orientation. The vertical lines overlaying the spectra correspond to energy level calculations of one possible crystal orientation. The heights of the peaks relate to the strength of the resonance. The spectra are off-set by 10\% for clarity b) Modification to the ODMR spectrum as polarisation orientation of the trapping laser is oriented through $90^\circ$. Clear peak shifts and contrast changes are observed. The vertical lines are predicted peak positions with a corresponding rotation of the crystal axis. Good agreement between theory and experiment is achieved. c) Graphical representation of the predicted coordination of the NV\textsuperscript{--} orientations relative to the external magnetic field. The ellipsoid represents the ND shape rotating with the polarisation axis and the coloured arrows represent the  NV\textsuperscript{--} orientations as suggested by the simulation of the eigenvalues for the system.}\label{fig:4}
\end{figure}

As a demonstration of the sensing capabilities we measure changes to the ODMR spectrum as a function of applied static magnetic field using an external rare earth magnet placed adjacent to the sample chamber (see Fig. \ref{fig:1}b). The magnet is oriented to give a significant in-plane magnetic field component and the strength is varied between 0 and 5 mT by adjusting the relative distance of the magnet (via the travelling stage) to the sample stage. The average field at the sample is calibrated against a Gauss meter. An example of the magnetic field dependence of the ODMR spectrum for a trapped ND is depicted in Fig.\,\ref{fig:4}a. With the magnet fully retracted the spectrum the ODMR spectrum (labelled) exhibits two strain-split peaks of width 12\,MHz, centred around 2.87\,GHz with a separation of 17 MHz and contrast of 10\%. With increasing B-field strength we observe an increase in the splitting of the ODMR peaks, a corresponding broadening of the resonance and a reduction in contrast. The splitting is associated with Zeeman splitting of the degenerate m\textsubscript{s} $ = \pm $ 1 levels of the NV\textsuperscript{--} ground and excited states, whilst the broadening and contrast reduction is associated with the splitting from different NV\textsuperscript{--} axis projections which result from the possible crystallographic orientations present in the trapped ND. As the Zeeman splitting energy depends on the dot product between the spin vector and magentic field, different NV\textsuperscript{--} orientations will experience different degrees of splitting. At small fields the splitting is less than the line widths of the individual resonances, but at the highest fields (i.e. 5\,mT) multiple bands are clearly observed with a reduced contrast ($\sim$ 3\%). 
\\
Immediately we can make the observation that the ND in this measurement must be constrained within the trap in terms of its orientation. As the resonance positions depend on the orientation of the NV\textsuperscript{--} axis with respect to the applied field and variation in orientation will yield an averaging of peak positions. In many experiments the ODMR spectra does not show clear resonances under high fields, but rather a broad featureless plateau of several hundred MHz with a low contrast ($<$ 1\%)  centred around the zero-field resonance position. In these cases the NDs do not maintain a fixed orientation within the trap, but are free to undergo Brownian angular diffusion over the duration of the measurements (i.e. hundreds of seconds). In this case the ODMR signal will be averaged over all possible NV\textsuperscript{--} axis. The variation in observed behaviour can be related to the variation of the size and shapes of individual NDs taken from samples prepared by ball mill processing \cite{Reineck2019}. 
\\
To confirm the orientation control of the trapped nanodiamond we record ODMR spectra from the same ND at an applied field of 5 mT with the trapping laser polarisation orientations set to $\phi = 0^\circ, 90^\circ$ and $45^\circ$, where $\phi$ is the azimuthal angle in the trapping plane. The results are presented in Fig.\,\ref{fig:4}b. Here we observe a clear shift in the peaks with a narrowing of the peak separation and a broadening and lowering of the individual bands. This indicates that the individual peaks associated with the NV\textsuperscript{--} axis projections onto the applied field have been modified.
\\
Calculating the energy spectrum for our estimated magnetic field configuration allows us to estimate the orientation of the NV\textsuperscript{--} unit cell in the lab frame. This is done by determining the eigenvalues of the NV\textsuperscript{--} Hamiltonian for each orientation class (denoted $j$), $H_j = D_\textrm{zfs}S_{z',j}^2 + \gamma\mathbf{B}\cdot \mathbf{S}_j$, for an arbitrarily-oriented magnetic field $\mathbf{B}$, where $z'$ is the N-V axis, $D_\textrm{zfs}=2.87$\,GHz is the zero field splitting energy, $\gamma = 28\,$GHz\,T$^{-1}$ is the electron gyromagnetic ratio, $\mathbf{S}$ is the spin vector, and for simplicity we have ignored additional effects of strain and electric fields in the model. We define a rotation operator $\mathcal{R}$ that describes the orientation of the NV unit cell, and project into the rotated NV frame the known lab-frame magnetic field orientation\cite{wood_t2-limited_2018}. Solving for the energy eigenvalues in the NV frame, we can approximate $\mathcal{R}$ and hence the orientation of the four NV orientation classes with respect to the magnetic field. Simulating $45^\circ$ stepped rotation of the nanodiamond about the $z$-axis, we can then generate spectra consistent with changes induced by the rotation of the trapping laser polarisation axis. Simulated resonance positions are presented in Fig.\,\ref{fig:4}a \& b as vertical lines where the height of each is proportional to the expected contrast. The lines between the spectra in Fig.\,\ref{fig:4}b trace the corresponding resonance shifts for the intervening angles. In Fig\,\ref{fig:4}c a three dimensional rendering of the predicted nanodiamond (ellipsoid) and NV\textsuperscript{--} orientations (coloured arrows) for the different trapping laser polarisation states. In the diagrams, the imaging / trapping plane is defined by the x-y axis and the z-axis is the laser propagation axis. The ellipse follows the polarisation direction of the trapping laser and NV\textsuperscript{--} orientations are fixed relative to the ND. We note that the simulations presented are one possible set of orientations that gives a satisfactory fit to the data, but we cannot preclude the possibility of other solutions having similar agreement. Further orientation dependent measurements would be necessary to narrow the possible solutions and also a refinement of the model that included broadening due to possible angular Brownian motion.
\\
Precise orientation control of the trapped ND is an essential feature of any trapped NV\textsuperscript{--} based vector magnetometry and enables a host of different types of applications, including those that utilise active control (e.g. spinning), such as quantum phase measurements \cite{wood_quantum_2018, wood_observation_2020}.  Preliminary demonstrations of orientation control for trapped NDs (containing a single NV\textsuperscript{--} centre) were presented by  \citeauthor{Geiselmann2013}. In these experiments the small applied field ($\sim$ 1.5 mT) and single projection limit the possibility of accurately determining the NV orientation completely. In the case of the bright NDs (containing many NV\textsuperscript{--} centres) the four possible NV projections give a much richer source of information to reliably reconstruct the field and orientation, as has been observed in other types of trapping technologies \cite{delord_electron_2017}. For optical trapping, the shape of a nanoparticle plays an important role in determining the equilibrium trapping orientation and dynamic behaviour and this has been exploited for microscopic objects to achieve specific functionality \cite{Simpson2014}. The ability to engineer specific shapes of nanodiamonds will be an important consideration for future application of trapped nanodiamond sensors and may play a dual purpose of optimising orientation control and emission properties for magnetic resonance sensing \cite{McCloskey:2020aa}.

\section{Conclusions}

In conclusion, we have demonstrated a method for achieving high quality optically detected magnetic resonance (ODMR) spectra from single nanodiamonds held in a gradient force optical tweezers. By temporarily modulating the trapping laser during optical spectroscopy measurements we are able to mitigate the deleterious effects of the infrared light on the fluorescence signal. Using this method we have demonstrated dual sensing (ODMR and T\textsubscript{1} measurements) together with precise Brownian motion measurements on sub-50 nm diamonds under strong trapping conditions. We also show DC magnetic field sensing capabilities of trapped 100 nm sized diamonds with an order of magnitude improvement sensitivity compared to the existing literature. Finally, we show that with increasing applied magnetic field we can resolve individual projections within the ODMR spectrum arising from different NV orientation classes aligned with a background field. By rotating the trapping laser polarisation we find that the resonances shifts in a manner consistent with rotation of the NV unit cell, as determined by a simple theoretical model. The new demonstrations represent important steps towards viable trapped-nanodiamond nanoscale vector-magnetometry and magnetic resonance sensing.

\bibliography{frontiers_LWR}


\end{document}